\documentclass[prl,aps,twocolumn]{revtex4}
\usepackage{graphicx}
\usepackage{dcolumn}
\usepackage{bm}

\def\beq{\begin{equation}}
\def\eeq{\end{equation}}
\def\beqa{\begin{eqnarray}}
\def\eeqa{\end{eqnarray}}

\newcommand{\roughly}[1]{\mathrel{\raise.3ex\hbox{$#1$\kern-0.85em
\lower1ex\hbox{$\sim$}}}}

\begin{document}

\preprint{APS/123-QED}

\title{Ultra-high-$Q$ TE/TM dual-polarized photonic crystal nanocavities}
\author{Yinan Zhang}
\email{yinan@seas.harvard.edu}
\author{Murray W. McCutcheon}
\author{Ian B. Burgess}
\author{Marko Loncar}
\affiliation{School of Engineering and Applied Sciences, Harvard University, Cambridge, MA 02138, USA}

\date{May 2009}

\begin{abstract}
We demonstrate photonic crystal nanobeam cavities that support both TE- and TM-polarized modes, each with a Quality factor greater than one million and a mode volume on the order of the cubic wavelength. We show that these orthogonally polarized modes have a tunable frequency separation and a high nonlinear spatial overlap. We expect these cavities to have a variety of applications in resonance-enhanced nonlinear optics.
\end{abstract}
\maketitle
Ultra-high Quality factor ($Q$) photonic crystal nanocavities, which are capable of storing photons within a cubic-wavelength-scale volume ($V_{mod}$), enable enhanced light-matter interactions, and therefore provide an attractive platform for cavity quantum electrodynamics \cite{yoshie, vuckovic} and nonlinear optics \cite{solj-rev, raineri, andreani, murray07, murray-lukin, DFG-OE, bravo-OE}. In most cases, high $Q/V_{mod}$ nanocavities are achieved with planar photonic crystal platform based on thin semiconductor slabs perforated with a lattice of holes. These structures favor transverse-electric-like (TE-like) polarized modes (the electric field in the central mirror plane of the photonic crystal slab is perpendicular to the air holes). In contrast, the transverse-magnetic-like (TM-like) polarized bandgap is favored in a lattice of high-aspect-ratio rods \cite{book, multipole}. TM-like cavities have been designed in an air-hole geometry, as well \cite{arakawa, triangular, painter}, but the $Q$ factors of these cavities were limited to the order of $10^3$. In addition, the lack of vertical confinement of these cavities results in large mode volumes \cite{arakawa}. Though it is possible to employ surface plasmons to localize the light tightly in the vertical direction, the lossy nature of metal limits the $Q$ to about $10^2$ \cite{painter}.

In this paper, we report a one-dimensional (1D) photonic crystal nanobeam cavity design that supports an ultra-high-$Q$ ($Q>10^6$) TM-like cavity mode with $V_{mod}\sim(\lambda/n)^3$. This cavity greatly broadens the applications of optical nanocavities. For example, it is well-suited for photonic crystal quantum cascade lasers, since the inter-subband transition in quantum cascade lasers is TM-polarized \cite{faist, forchel, loncar-QCL}. We also demonstrate that our cavity simultaneously supports two ultra-high-$Q$ modes with orthogonal polarizations (one TE-like and one TM-like). The frequency difference of the two modes can be widely tuned while maintaining the high $Q$ factor of each mode, which is of interest for applications in nonlinear optics.

\begin{figure}[htbp]
\centering
\includegraphics[width=7.8cm]{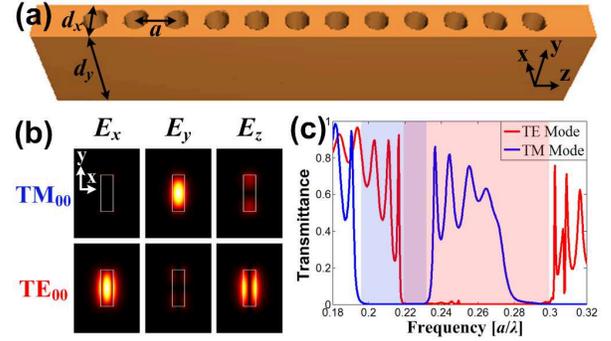}
\caption{\label{F1} (a) Schematic of the nanobeam design, showing the nanobeam thickness ($d_{y}$) and width ($d_{x}$), and the hole spacing ($a$). (b) TE$_{00}$ and TM$_{00}$ transverse mode profiles for a ridge waveguide with $d_{y}=3d_{x}$. (c) Transmission spectra for the TE$_{00}$ (red) and TM$_{00}$ (blue) Bloch modes. The shaded areas indicate the bandgaps for both modes.}
\end{figure}

Our design is based on a dielectric suspended ridge waveguide with an array of uniform holes of periodicity, $a$, and radius, $R$, which form a 1D photonic crystal Bragg mirror \cite{foresi}, as shown in Fig. 1(a). The refractive index of the dielectric is set to $n=3.4$ (similar to Si and GaAs at $\sim$ $1.5\mu m$). We first start with a ridge of height:width:period ratio of 3:1:1 ($d_{x}=a$, $d_{y}=3a$) and $R=0.3a$. Fig. 1(b) shows the transverse profiles of the fundamental TM-like and TE-like modes (TM$_{00}$ and TE$_{00}$) supported by the ridge waveguide. The TM$_{00}$ mode has its major component ($E_{y}$) lined along the hole axis, whereas the TE$_{00}$ mode's major component ($E_{x}$) is perpendicular to the air holes. Using the three-dimensional (3D) finite-difference time-domain (FDTD) method, the transmittance spectra are obtained of the TM$_{00}$ and TE$_{00}$ modes launched towards the Bragg mirror. Fig. 1(c) shows the TM$_{00}$ and TE$_{00}$ bandgaps, respectively. In contrast to two-dimensional (2D) photonic crystal slabs, where the photon is localized in the $xz$ plane via Bragg scattering, here we only require Bragg confinement in the longitudinal ($z$) direction, as light is transversely confined in the other two dimensions by total internal reflection. It has also been shown experimentally that 1D photonic crystal nanobeam cavities have comparable $Q/V_{mod}$ ratios to 2D systems \cite{parag, delarue}.

\begin{figure}[htbp]
\centering
\includegraphics[width=6.6cm]{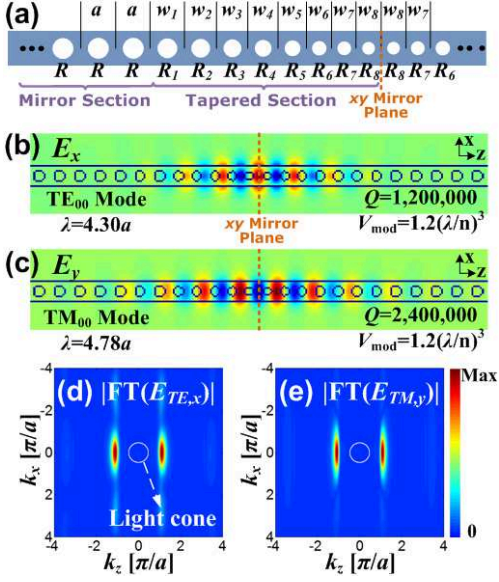}
\caption{\label{F2} (a) Schematic of the 1D photonic crystal nanobeam cavity, with the tuning parameters $R_{k}$ and $w_{k}$ in the 8-segment tapered design. (b,c) Mode profiles of the electric field components $E_{TE,x}$ and $E_{TM,y}$ for the cavity design with $d_{x}=a$, $d_{y}=3a$. (d,e) Spatial Fourier transform of the electric field component profiles ($E_{TE,x}$ and $E_{TM,y}$) in the $xz$ plane ($y=0$).}
\end{figure}

Introducing a lattice grading to the periodic structure creates a localized potential for both TE- and TM-like modes. To optimize the mode $Q$ factors, we apply the bandgap-tapering technique that is well-developed in previous work \cite{yinan-pillar,yinan-wires,murraySiN,notomi,zipper}. We use an 8-segment tapered section with holes ($R_{1}$-$R_{8}$) and a 12-period mirror section at each side. Two degrees of freedom are available for each tapered segment: the length ($w_{k}$) and the radius ($R_{k}$). We keep the ratio $R_{k}/w_{k}$ fixed at each segment, and then implement a linear interpolation of the grating constant ($2\pi/w_{k}$). When the central segment $w_{8}$ is set to $0.84a$, we obtain ultra-high $Q$s and low mode volumes for both TE- and TM-polarized modes ($Q_{TE}=1.2\times 10^6$, $Q_{TM}=2.4\times 10^6$, $V_{mod,TE}=V_{mod,TM}=1.2(\lambda/n)^3$), with free-space wavelengths $4.30a$ and $4.78a$, respectively. Fig. 2(b) and (c) show the mode profiles of the major components of the two modes in the {\it xz} mirror plane. The ultra-high $Q$ factors can also be interpreted in momentum space \cite{momentum_I, momentum_II, arakawa}. Fig. 2(d) and (e) demonstrate the Fourier transformed (FT) profiles of the electric field components $E_{TE,x}$ and $E_{TM,y}$ in the {\it xz} plane ($y=0$), with the light cone indicated by the white circle. It can be seen that both modes' Fourier components are localized tightly at the bandedge of the Brillouin zone on the $k_{z}$-axis ($k_z=\pi/a$). This reduces the amount of mode energy within the light cone that is responsible for scattering losses. It is also worthwhile to note that higher-longitudinal-order TE$_{00}$ and TM$_{00}$ cavity modes with different symmetry with respect to the $xy$ mirror plane exist \cite{yinan-pillar}. For example, the second-order TE$_{00}$ mode, which has a node at the $xy$ mirror plane, resonates at a wavelength of $4.43a$. It has a higher $Q$ factor of $4.7 \times 10^6$, but a larger mode volume of $2.1(\lambda/n)^3$.

For a number of applications of interest, control of the frequency spacing between the two modes is required. Examples include polarization-entangled photon generation for degenerate modes \cite{imamoglu}, and terahertz generation for $0.1-10$THz mode splitting \cite{DFG-OE}. We tune the frequency separation of the two modes by varying the thickness of the structure while keeping the other parameters constant. In Fig. 3(a), the cavity resonances of the TE$_{00}$ and TM$_{00}$ modes are traced as a function of the nanobeam thickness ($d_{y}/a$). The TM-like modes have a much larger dependence on the thickness than the TE-like modes. The modes are degenerate at $d_{y}=1.26a$, and for thicknesses beyond this value, $\omega_{TE}$ is larger than $\omega_{TM}$. As $d_y$ increases, the splitting increases until it saturates when the system approaches the 2D limit (structure is infinite in the {\it y}-direction). In this limit, we find that $\lambda_{TE}=4.4a$ and $\lambda_{TM}=5.1a$. The frequency separation ($\delta\omega=|\omega_{TE}-\omega_{TM}|$) of this design ranges from 0THz to 20THz, with the TE-like mode wavelength fixed at $1.5\mu m$ by scaling the structure accordingly. Fig. 3(b) shows the thickness dependence of the $Q$ factor for the {\it xz} design specifications listed above. It can be seen that the $Q$ factors of both TE- and TM-polarized modes stay above $10^5$ for the $\omega_{TE}>\omega_{TM}$ branch.

\begin{figure}[htbp]
\centering
\includegraphics[width=8.3cm]{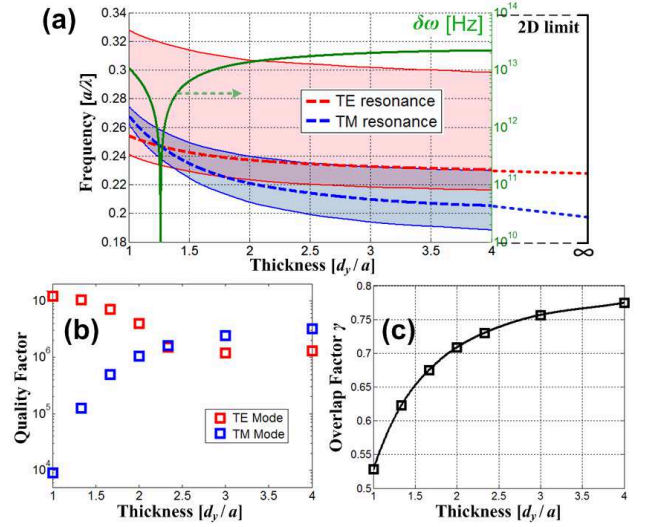}
\caption{\label{F3} (a) TE$_{00}$ (red) and TM$_{00}$ (blue) cavity mode resonant frequencies (dotted lines) as a function of the nanobeam thickness. The bandgap regions of the two modes are shaded. The frequency separation ($\delta\omega$) of the two modes with the TE-like mode wavelength fixed at $1.5\mu m$ by scaling the structure accordingly is plotted in green. (b,c) Dependence of the $Q$ factor and nonlinear overlap factor $\gamma$ on the nanobeam thickness.}
\end{figure}

Decreasing $d_y$ causes the width of the TM bandgap to sharply decrease, whereas the width of the TE bandgap remains almost constant. The narrowed TM bandgap results in a reduced Bragg confinement, which increases the transmission losses through the Bragg mirrors. This is evidenced by the $Q$ factor of the TM mode, which drops to 9,000 when the thickness:width ratio is 1:1. Though this leakage can be compensated for, in principle, by increasing the number of periods of the mirror sections, the length of the structure also increases, which makes fabrication more challenging for a suspended nanobeam geometry. A narrow bandgap also leads to large penetration depth of the mode into the Bragg mirrors, thereby increasing the mode volume.

Next, we examine the application of our dual-polarized cavity for the resonance enhancement of nonlinear processes. To achieve a large nonlinear interaction in materials with dominant off-diagonal nonlinear susceptibility terms (e.g. $\chi_{ijk}^{(2)}, i\neq j\neq k$), such as III-V semiconductors \cite{murray07, murray-lukin, GaAs-ref}, it is beneficial to mix two modes with orthogonal polarizations. As shown in our previous work ~\cite{DFG-OE}, the strength of the nonlinear interaction can be characterized by the modal overlap, which can be quantified using the following figure of merit,
\beq
\gamma\equiv\epsilon_{r,d}\frac{\int_{d} d^3{\bf r}\sum_{i,j,i\neq j}E_{TE,i}E_{TM,j}}{\sqrt{\int d^3{\bf r}\epsilon_{r}|{\bf E}_{TE}|^2}\sqrt{\int d^3{\bf r}\epsilon_{r}|{\bf E}_{TM}|^2}}.
\eeq
where $\int_{d}$ denotes integration over only the regions of nonlinear dielectric, and $\epsilon_{r,d}$ denotes the maximum dielectric constant of the nonlinear material. Note that we have normalized $\gamma$ so that $\gamma=1$ corresponds to the theoretical maximum overlap. For the TE$_{00}$ and TM$_{00}$ modes we studied, the two major components ($E_{TE,x}$ and $E_{TM,y}$) share the same parity (have anti-nodes in all the three mirror planes), and only two overlap components, $E_{TE,x}E_{TM,y}$ and $E_{TE,y}E_{TM,x}$, in Eq. (1) do not vanish. This allows a large nonlinear spatial overlap. We obtain $\gamma=0.76$ for the cavity shown in Fig. 2. The overlap approaches $\gamma=0.78$ in the limit $d_{y}\rightarrow\infty$. We find that the overlap factor, $\gamma$, stays at a reasonably high value ($>0.6$) across the full range of the frequency difference tuning (for $\omega_{TE}>\omega_{TM}$ branch) [Fig. 3(c)].

\begin{figure}[htbp]
\centering
\includegraphics[width=7cm]{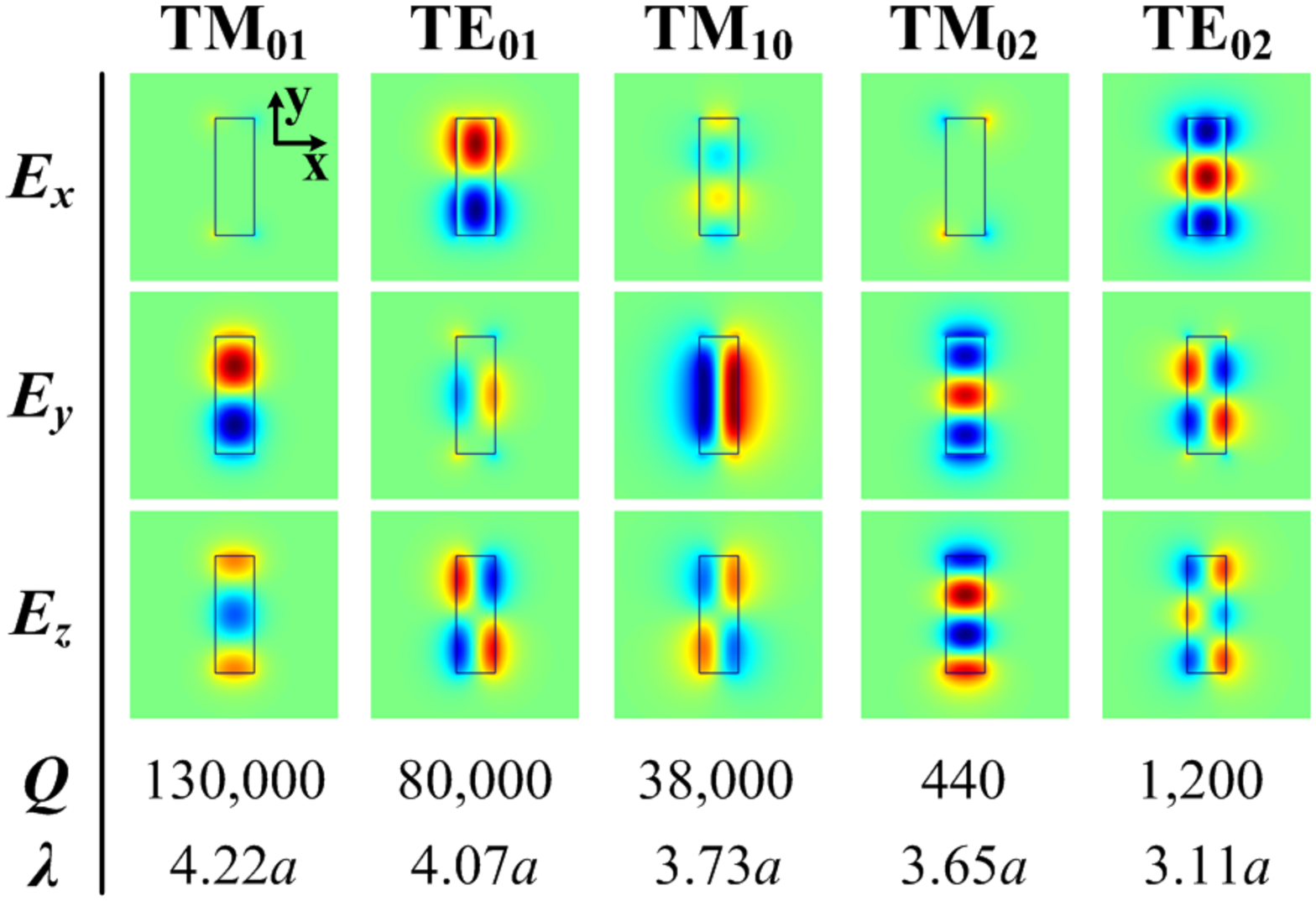}
\caption{\label{F4} Parameters of the higher-order cavity modes for the design with $d_{x}=a$, $d_{y}=3a$.}
\end{figure}

Finally, it is important to note that thick nanobeams can support higher-order modes with a different number of nodes in the $xy$ plane, as well. These higher-order modes are also confined in the tapered section within their respective bandgaps, with the $Q$ factors and wavelengths listed in Fig. 4 for the $d_{x}=a$ and $d_{y}=3a$ case. These modes can offer a broader spectral range than the fundamental modes, which is of great interest to nonlinear applications requiring a large bandwidth \cite{murray-lukin}.

In conclusion, we have demonstrated that ultra-high-$Q$ TE- and TM-like fundamental modes with mode-volumes $\sim (\lambda/n)^3$ can be designed in 1D photonic crystal nanobeam cavities. We have shown that the frequency splitting of these two modes can be tuned over a wide range without compromising the $Q$ factors. We have also shown that these modes can have a high nonlinear overlap in materials with large off-diagonal nonlinear susceptibility terms across the entire tuning range of the frequency spacing. We expect these cavities to have broad applications in the enhancement of nonlinear processes.

This work was supported in part by National Science Foundation (NSF) and NSF career award. M.W.M and I.B.B wish to acknowledge NSERC (Canada) for support from PDF and PGS-M fellowships.

\end{document}